\documentclass[journal]{IEEEtran}

\usepackage{cite}
\usepackage{color}
\usepackage{graphicx}
\usepackage{amsmath}
\usepackage{array}
\usepackage{url}
\usepackage{braket}

\begin{document}
%
\title{Quantum Biomimetic Modeling of Diamond NV$^{-}$ Center Spin Dynamics}

\author{Vishvendra~Singh~Poonia,
        Dipankar~Saha,
        and~Swaroop~Ganguly,~\IEEEmembership{Member,~IEEE}%
        
\thanks{V. S. Poonia, D. Saha and S. Ganguly are with the Department
of Electrical Engineering, Indian Institute of Technology Bombay, Mumbai, 
400076 India. e-mail: sganguly@ee.iitb.ac.in; vishvendra.punia@gmail.com.}%
}
\vspace*{\fill}
\LARGE {This work has been submitted to the IEEE for possible publication. Copyright may be transferred without notice, after which this version may no longer be accessible.}
\vspace*{\fill}

\maketitle

\begin{abstract}
The spin dynamics of the diamond NV$^{-}$ center turns out to be similar to that of the chemical compass responsible for avian magnetoreception. We present a simulation framework based on quantum master equation for the former that is based upon the Radical Pair model of the latter. We show that this framework captures all the experimentally studied behavior of the NV$^{-}$ center spin system and can therefore be a predictive modeling tool.
\end{abstract}


\IEEEpeerreviewmaketitle

\normalsize
\section{Introduction}
\IEEEPARstart{F}{ew-spin} systems are rich playgrounds for quantum phenomena in myriad contexts from quantum information sciences to quantum biological systems, and have therefore generated intense research interest. In particular, the NV$^{-}$ color center in diamond, on account of its unique electronic and optical properties, has been extensively explored as a candidate for solid-state quantum computing, quantum metrology and sensing applications \cite{doherty2013nitrogen,balasubramanian2008nanoscale,cai2014hybrid}. Diamond NV$^{-}$ center based magnetometry can be performed with nanoscale spatial resolution, making it an ideal system for imaging weak magnetic fields with high resolution and for ~\cite{doherty2013nitrogen}.

The nitrogen-vacancy (NV) center is a naturally occurring defect in the diamond crystal which comprises a substitutional nitrogen atom adjacent to a vacancy in the diamond lattice. Two kind of NV defects occur naturally in diamond lattice: charge-neutral (denoted as NV$^{0}$) and negatively-charged (denoted as NV$^{-}$). The NV$^{-}$ state is highly photostable and lends itself to high-precision magnetometry experiments \cite{balasubramanian2009ultralong,bar2013solid}. In this work, we deal only with the negatively charged NV defect. Three key characteristics of the NV$^{-}$ center makes it a suitable candidate for magnetometry and other applications, viz. its unique electronic structure, optical polarization of the center and optical detection of its spin states. Its electronic structure consists of ground and excited spin-triplet states ($^{3}A_2$ and $^{3}E$ respectively) with two intermediate singlet states ($^{1}E$ and $^{1}A_1$), all situated in the bandgap of the diamond~\cite{acosta2010optical}.
The intermediate singlet states play an important role in the overall spin dynamics here. The transitions between the $^{3}A_2$ and the $^{3}E$ states are radiative. However, the $m_s = \pm 1$ sublevels ($m_s$ here denotes the electron-pair spin state) of the $^{3}E$ state can also have non-radiative relaxation to the singlet states, and thereafter into the $m_s = 0$ sublevel of the $^{3}A_2$ state. Under optical pumping (from ground to excited state), this leads to a net population transfer from the $m_s = \pm 1$ sublevels into the $m_s = 0$ sublevel of the ground triplet state -- a process called optical polarization. Additionally, the spin sublevels of the ground state has different fluorescence for $m_s = 0$ and $m_s = \pm 1$ that makes the detection of spin states possible by optical means~\cite{balasubramanian2008nanoscale}.

\begin{figure}[!t]
  \centering
      \includegraphics[width=2.5in]{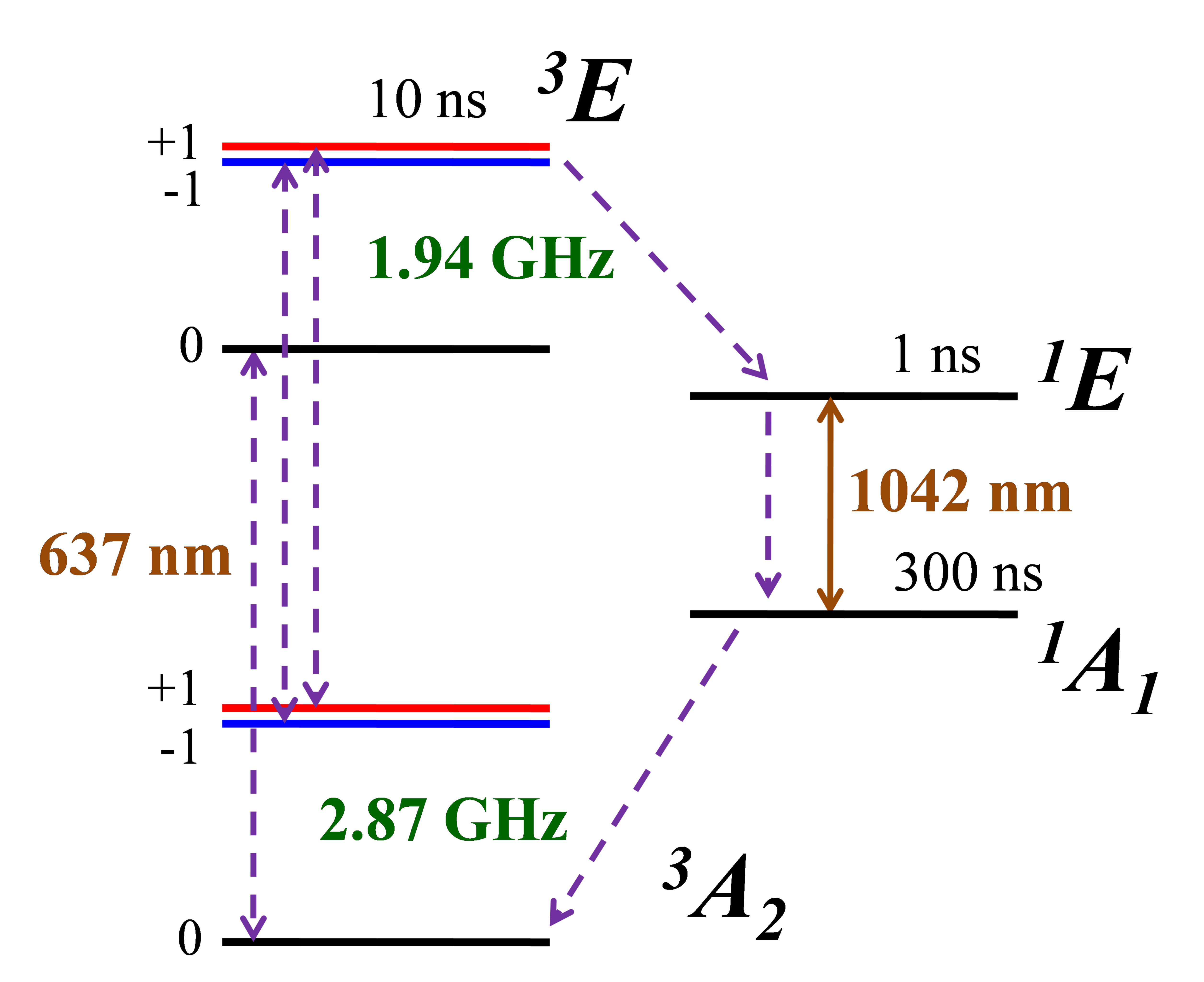}
      \caption{Diamond $NV^{-}$ center energy level diagram. Energy of various sublevels of ground and excited states, along with intermediate singlet states, is shown herein. The timescales shown corresponding to the levels are the typical lifetimes of these energy levels.}
  \label{Fig1_EnergyLevels}
\end{figure}

The aforementioned transitions, along with the typical lifetimes of the states, are illustrated in Fig.~\ref{Fig1_EnergyLevels}. In addition to optical polarization, the diamond NV$^{-}$ center also exhibits long coherence time for the ground state~\cite{balasubramanian2009ultralong,bar2013solid}. Optical polarization and long coherence time of diamond NV$^{-}$ center makes it an ideal candidate for high-precision nanoscale magnetometry.

In this work, we present a quantum master equation based simulation framework to model in a unified way the spin dynamics and optical transitions of the diamond NV$^{-}$ center for quantum information and quantum metrology applications. We call this approach quantum biomimetic in the sense that it is inspired by a quantum biological system, i.e. a biological system in which quantum mechanics plays a non-trivial functional role~\cite{lambert2013quantum}. Specifically, it is akin to the simulation framework of radical pair (RP) spin dynamics of the avian compass~\cite{gauger2011sustained,poonia2015state}, viz. the ability of certain migratory bird species to sense the geomagnetic field and use it in navigation \cite{schulten1978biomagnetic,ritz2000model,poonia2017functional,poonia2015state}. The RP models the spin dynamics of a radical pair generated by photons in an eye of the migratory bird. Each radical hosts an unpaired electron spin. One of these electrons is devoid of nuclear hyperfine interaction, while the other feels a strong hyperfine coupling; the geomagnetic (Zeeman) field, on the other hand, affects both the spins in a similar manner. The Hamiltonian governing the spin dynamics is given as~\cite{gauger2011sustained}:
\begin{eqnarray}
\label{Eq1_AvianHamiltonian}
H =\gamma \mathbf{B} \cdot (\hat{S_1} + \hat{S_2}) + \hat{I} \cdot \mathbf{A} \cdot\hat{S_2}
\end{eqnarray}
$\hat{S_1}$, $\hat{S_2}$, and $\hat{I}$ are the electronic and nuclear spin operators respectively. $\mathbf{A}$ denotes the hyperfine interaction tensor between second electron and its nearby nucleus. The first term on the right hand side models the Zeeman interaction of the radical pair spins with the local geomagnetic field, while the second term models the hyperfine interaction between one of those spins and the proximate nucleus. The spin state the radicals can either be singlet or triplet or a superposition thereof before they recombine; the recombination product coming from the singlet states (known as the singlet yield) depends on the inclination of the Zeeman field (local geomagnetic field) which results in the compass operation. In recent work, we solved the radical pair spin dynamics using Lindblad type quantum master equation, wherein Lindblad operators modeled the recombination process and decoherence due to environmental interaction~\cite{poonia2015state}.

Here, we generalize the framework proposed in \cite{gauger2011sustained,poonia2015state} to model the spin dynamics of NV$^{-}$ center. Notably, the NV$^{-}$ center Hamiltonian is strikingly similar to that of the RP model of avian compass, which is essentially the reason that it was earlier proposed as a potential candidate biomimetic system to mimic the mechanism of avian compass~\cite{cai2010quantum}. 
In fact, the optical transitions in the NV$^{-}$ center can indeed be modeled in the same way as radical pair recombination through the singlet and triplet channels in the avian compass. We have validated our quantum biomimetic simulation framework by modeling the results of available experimental studies on the diamond NV$^{-}$ center~\cite{poonia2016quantum}. Thus, it can now be used as a design and modeling tool for magnetometry or quantum information processing devices based on the diamond NV$^{-}$ center, or similar spin systems.
\section{Model Description}
The ground and excited state Hamiltonians of the NV$^{-}$ center are given by \cite{doherty2013nitrogen}:
\begin{eqnarray}
\begin{split}
H_{gs} = D_{gs}[\hat{S_z}^2 - \frac{2}{3}I] + \hat{S} \cdot \mathbf{A} \cdot\hat{I} + \gamma \mathbf{B} \cdot \hat{S} \\
+ P_{gs}[\hat{I_z}^2 - \frac{2}{3}I]
\end{split}
\label{Eq_NVGroundHamiltonian}
\end{eqnarray}
\begin{eqnarray}
\begin{split}
H_{es} = D_{es} [\hat{S_z}^2 - \frac{2}{3}I] + \hat{S} \cdot \mathbf{A} \cdot\hat{I} + \gamma \mathbf{B} \cdot \hat{S} \\
+ P_{es} [\hat{I_z}^2 - \frac{2}{3}I]
\end{split}
\label{Eq_NVExcitedHamiltonian}
\end{eqnarray}
In these equations, the first term on the right-hand side (RHS) is the zero-field splitting term and arises from the spin-spin interaction of the vacancy electrons. The second and the third terms on the right-hand side are respectively the hyperfine and Zeeman interaction terms, while the fourth term is the nuclear self-Hamiltonian term. The nuclear term is negligible in comparison to the ZFS term and can be ignored generally. Effectively, this system comprises of a spin-1 system (two electron spins in the vacancy) interacting with the $^{14}$N nuclear spin. The similarity with the RP Hamiltonian (Eq.1) is notable here with the NV$^{-}$ center system differing only in the crystal-field and nuclear-spin Hamiltonian terms. Here we formulate a simulation scheme for the NV$^{-}$ system along the lines of the RP simulation framework~\cite{gauger2011sustained,poonia2015state} . Since a spin-1 system (vacancy spins) interacts with another spin-1 system ($^{14}$N nucleus), the Hilbert spaces of ground and excited states are 9-dimensional. Apart from the spin dynamics in ground and excited states, optical transitions are responsible for transferring the spin populations between these states. In order to account for the optical transitions and spin dynamics concurrently, we construct a 20-dimensional Hilbert space. The first 9-dimensional subspace of this Hilbert space accounts for the ground state spin dynamics, the next 9-dimensional subspace accounts for the excited state spin dynamics, and the remaining 2-dimensional subspace accounts for the intermediate singlet states. Transitions between these are modeled by inter-subspace projection operators. The spin dynamics of the diamond NV$^{-}$ center can, thus, be modeled by the Lindblad type quantum master equation, given by:
\begin{eqnarray}
\label{Eq4_ME}
\dot \rho = - \frac{i}{\hbar}[H, \rho] + \sum\limits_{i=1}^{19} k_i (P_i \rho P_i^\dagger - \frac{1}{2}(P_i^\dagger P_i \rho + \rho P_i^\dagger P_i))
\end{eqnarray}
\begin{figure}[htbp]
 \centering
\includegraphics[width=8.5cm,keepaspectratio]{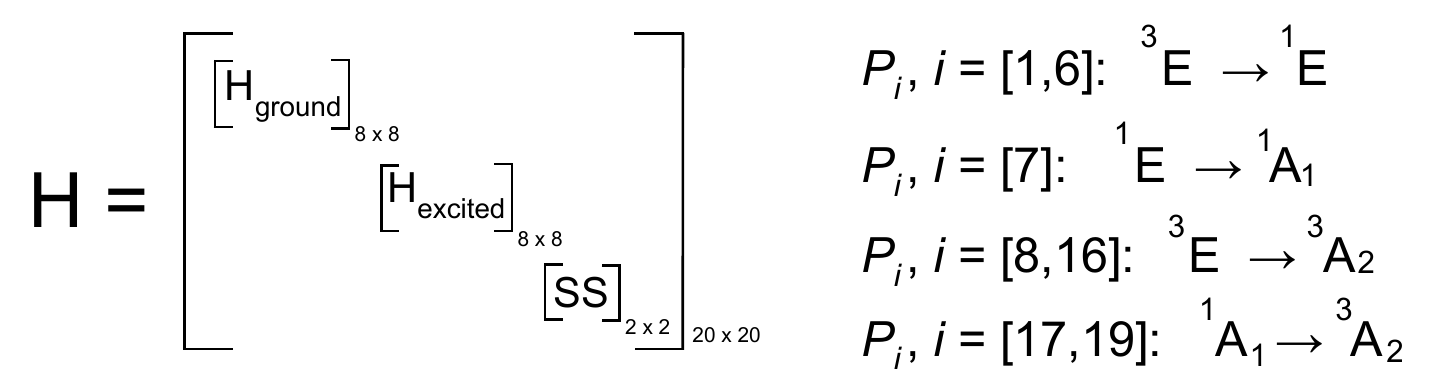}
      \caption{Scheme for constructing the Hamiltonian for modeling the dynamics of diamond NV$^{-}$ center. The Hamiltonians of ground state, excited state and intermediate singlet states are modeled by different subspaces of the composite Hamiltonian. The optical and spin dynamical transitions among the spin states are modeled using projection operators.}
      \label{Fig2_JointHamiltonian}
\end{figure}
The first term on the right-hand side is the unitary evolution term, describing the evolution of spin states under the Hamiltonian -- without the other terms on the RHS, this would be the von Neumann Equation for the density matrix. The second term on the RHS accounts for the transitions from excited state to the ground state via various pathways. Mathematically, these transitions are modeled using the projection operators viz. $P_i , i = [1,19]$. The projection operators $P_i: i = [1,6]$ correspond to the transitions from $m_s = \pm1$ and $m_I = 0, \pm1$ of $^{3}E$ state to $^{1}E$ state where $m_I$ represents the nuclear spin state. $P_7$ accounts for the transition from $^{1}E$ to $^{1}A_1$ state. Similarly, $P_i: i = [8,16]$ account for the transitions from $m_s = 0, \pm1$ and $m_I = 0, \pm1$ of $^{3}E$ state to $m_s = 0, \pm1$ and $m_I = 0, \pm1$ of the $^{3}A_2$ ground state and $P_i: i = [17,19]$ model the transition from $^{1}A_1$ state to $m_s = 0$ and $m_I = 0, \pm1$ of the $^{3}A_2$ ground state. $k_i$ corresponding to these states are the inverse of lifetimes of these states which are shown in Fig.~\ref{Fig1_EnergyLevels}. Additional details about the quantum master equation presented here is given in Appendix~\ref{Appendix1_QME}. The initial state of the center is assumed to be the excited triplet state i.e. the state after initial laser excitation. The ensuing evolution is modeled using this master equation. The final output is measured in the form of populations of ground state spin-sublevels. We use the Quantum Toolbox in Python (QuTiP) for simulating the master equation~\cite{johansson2012qutip}. The Hamiltonian and the projection operators are elaborated in Fig.~\ref{Fig2_JointHamiltonian}. \\
\begin{figure}[b]
      \centering
      \includegraphics[width=8.5cm,height=8.5cm,keepaspectratio]{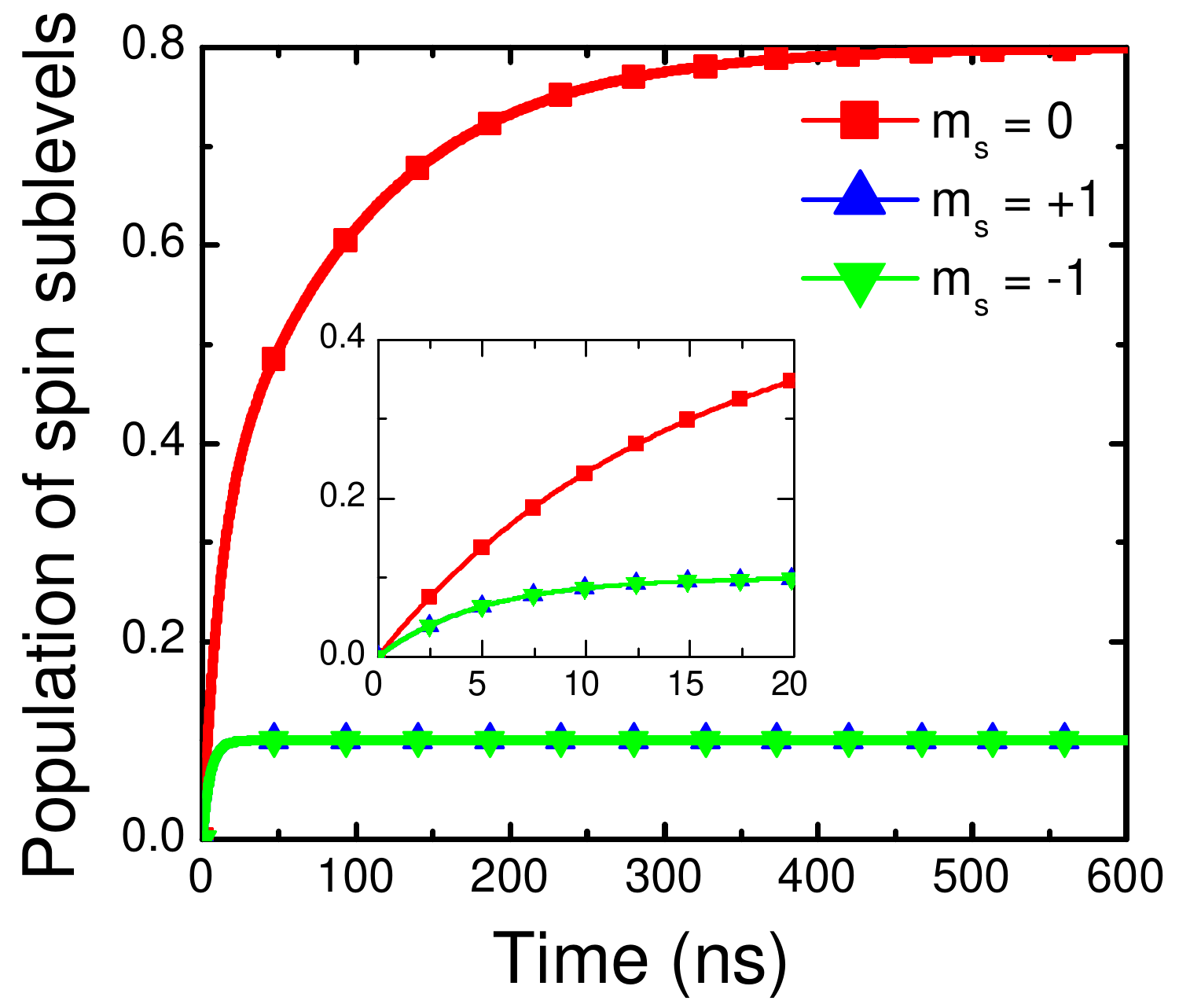}
      \caption{Population of ground state spin sublevels is plotted as a function of time after initial excitation by 532 nm green laser. The figure illustrates the modeling of optical polarization of the NV$^{-}$ center ground state.}
      \label{Fig3_OpticalPolarization}
\end{figure}
\section{Results}
In this section, we simulate the key characteristics of the diamond NV$^{-}$ center and demonstrate the modeling of  magnetometry experiments in this system. We start with the optical polarization after the NV$^{-}$ center spin is excited. Then we simulate NV$^{-}$ based magnetometry, both ODMR based DC magnetometry and AC magnetometry.

\subsection{Optical Polarization}
After the application of a 532 nm laser, the NV$^{-}$ center gets initialized in the $^{3}E$ state. We take this state as the initial state of the system. The following dynamics of the NV$^{-}$ center is simulated using the master equation of Eq.~\ref{Eq4_ME}. The populations of the ground state spin sublevels are calculated and plotted as a function of time in Fig.~\ref{Fig3_OpticalPolarization}. The plot illustrates that the population of $m_s=0$ spin sublevel reaches 80\% of the total population after around 450 ns. This kind of selective filling of $m_s = 0$ sublevel is termed as `polarization'. Non-radiative transitions from the $^{3}E$ $m_s = \pm1$ spin-sublevels to the singlet system and subsequently to the $^{3}A_2$ $m_s= 0$ sublevels, as illustrated in Fig.~\ref{Fig1_EnergyLevels}, play key role in this process. The timescale of polarization is governed by the largest lifetime of the intermediate states, viz. that of the $^{1}A_1$ singlet state. The populations of $m_s= \pm1$ spin sublevels saturate in 10-20 ns, which correspond to the timescale for radiative transition from the $^{3}E$ $m_s= \pm1$ sublevels to the corresponding $^{3}A_2$ sublevels. Notably, these timescales are consistent with the experimentally observed values~\cite{acosta2010optical}. Thus the simulation framework proposed herein successfully models the optical polarization of the NV$^{-}$ center -- which also lays the foundation for diamond NV$^{-}$ center based magnetometry. 
\begin{figure}[t]
      \centering
      \includegraphics[width=8.5cm,height=8.5cm,keepaspectratio]{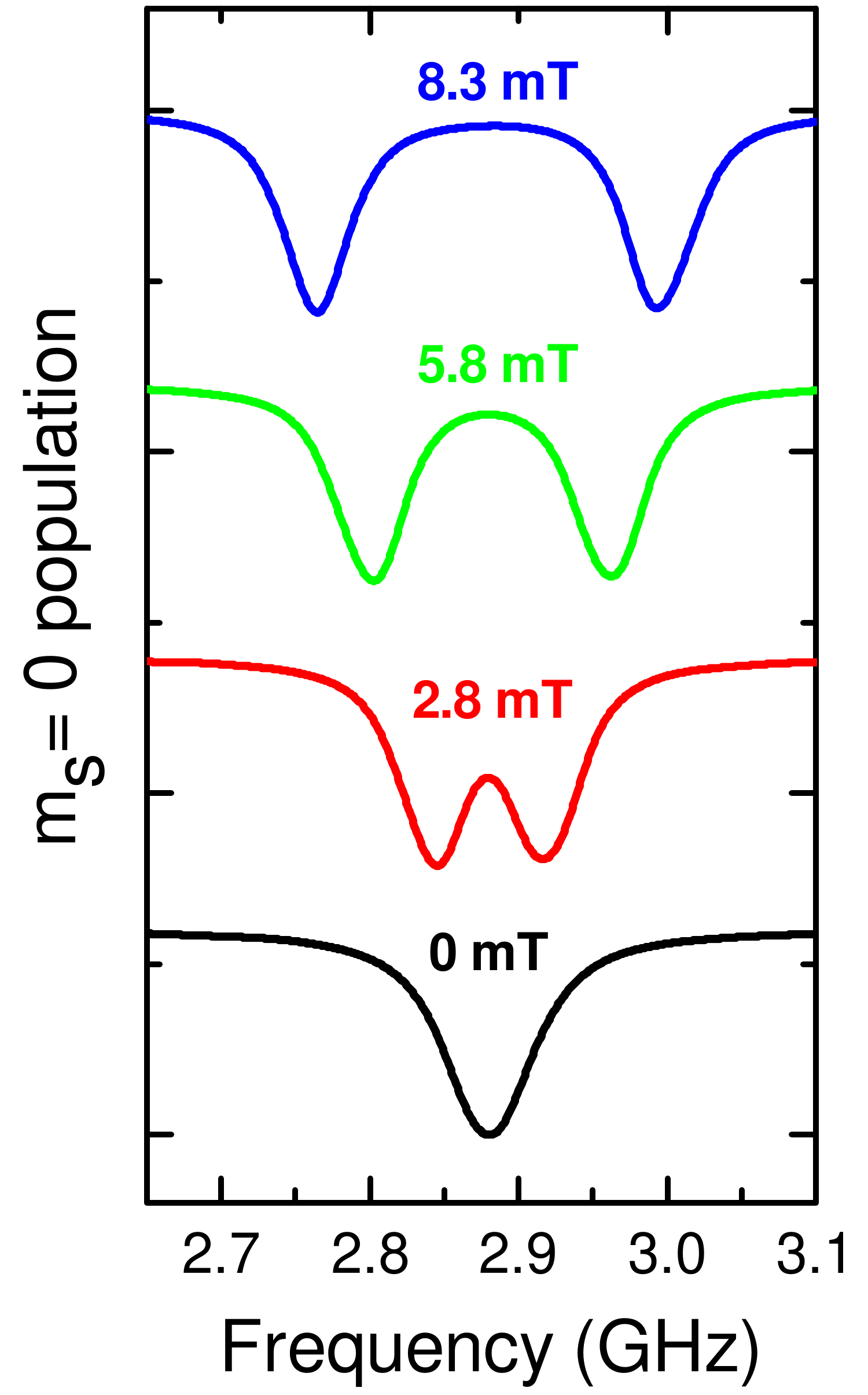}
      \caption{Scheme of detecting the dc magnetic fields using optically detected magnetic resonance (ODMR). DC field causes splitting of $m_s = \pm1$ levels that is detected by applying microwave fields to induce transitions between $m_s=0$ and $m_s=\pm1$. The transitions cause the population of $m_s = 0$ sublevel (hence the fluorescence of the center~\cite{balasubramanian2008nanoscale}) to dip and the distance between these dips is an indication of the magnitude of the applied DC magnetic field.}
      \label{Fig4_ODMRDCMag}
\end{figure}

\subsection{Magnetometry}
In this section, we introduce the elements of diamond NV$^{-}$ center based magnetometry as a preface to the results of our work. Magnetometry, in this context, usually involves a single NV$^{-}$ center as a scanning probe or ensemble of NV$^{-}$ centers for magnetic imaging~\cite{balasubramanian2008nanoscale,taylor2008high}. Here we describe single NV$^{-}$ center based magnetometry wherein the ground triplet state plays a pivotal role. Its long coherence time and accessibility to optical detection makes it especially  suitable for magnetometry applications. The Hamiltonian of the ground state is given by Eq.~\ref{Eq_NVGroundHamiltonian}. The first term in this equation accounts for the zero field splitting (ZFS) of the $m_s = 0$ and $m_s = \pm 1$ sublevels due to crystal field. The third term corresponds to the Zeeman interaction that causes splitting of the $m_s = \pm 1$ sublevels. This splitting, being $g \mu_B B$, provides a direct measure of the magnitude of the magnetic field applied along the axis connecting nitrogen to the vacancy. It can be detected using a microwave field and exploiting the differential fluorescence of $m_s = 0$ sublevel versus the $m_s = \pm 1$ sublevels~\cite{balasubramanian2008nanoscale,rondin2014magnetometry}. A dip in the fluorescence is seen at the microwave frequency that induces a transition between the $m_s = 0$ and $m_s = \pm 1$ sublevels. In the presence of a Zeeman field, there are two dips, one corresponding to the $m_s = 0 \leftrightarrow m_s = +1$ transistion, and the other corresponding to the $m_s = 0 \leftrightarrow m_s = -1$ transition. The spacing between these peaks immediately gives the magnitude of the applied magnetic field. This DC magnetometry technique is called Optically Detected Magnetic Resonance (ODMR).
In our proposed framework, the resonant microwave field part is included in the Zeeman term of the Hamiltonian in eq.~\ref{Eq_NVGroundHamiltonian} and spin sublevels' population is calculated as a function of the frequency of the microwave field. 
The population of ground state $m_s = 0$ sublevel as a function of the frequency of the applied microwave field for various DC magnetic field strengths is shown in Fig.~\ref{Fig4_ODMRDCMag}. We observe the expected dips in the $m_s = 0$ population and the Zeeman splitting of the dips. Here, the dip width obviously sets the limit on the sensitivity of diamond-based DC magnetometer to $\sim$1 mT \cite{degen2008scanning}. Presence of environmental noise widens these peaks and thus deteriorates the sensitivity of the magnetometer.

\begin{figure}[htbp]
 \centering
\includegraphics[width=8.5cm,height=8.5cm,keepaspectratio]{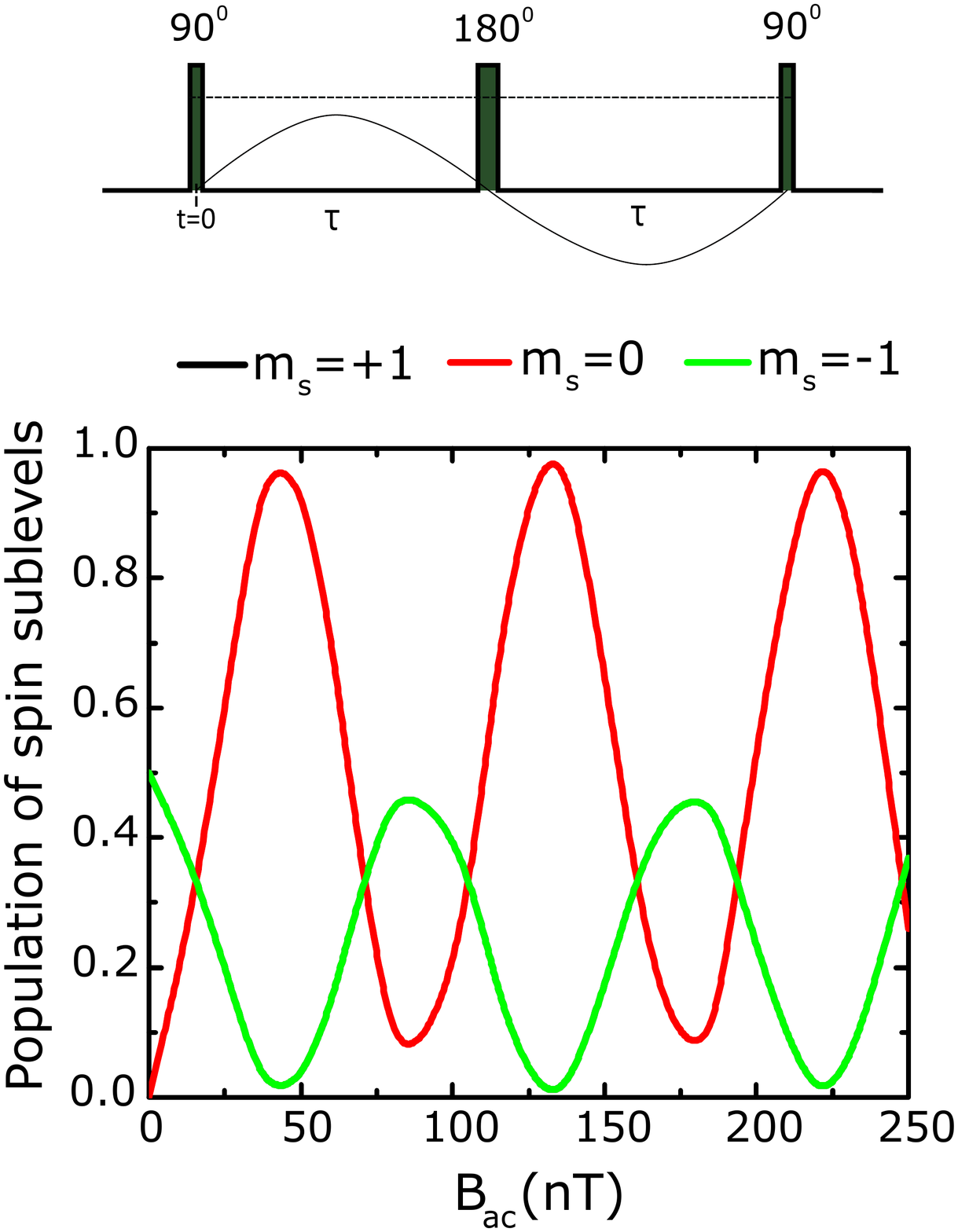}
      \caption{Detection of ac magnetic fields is demonstrated here along with the pulse sequence used. The $180^{\circ}$ pulse cancels any dc component of the magnetic field or noise and allows only the ac component of the magnetic field to get accumulated in the phase of the spin states. The cosine behavior of the $m_s = 0$ population captures the ac magnetic field amplitude~\cite{maze2008nanoscale}.}
      \label{Fig5_ACMag}
\end{figure}

The ODMR technique is suitable for detecting DC magnetic fields. In order to detect AC magnetic fields with better sensitivity, specific pulse sequences need to be designed and applied that can capture the field strength information. In this scheme, the diamond NV$^{-}$ center is optically polarized in $m_s = 0$ sublevel of ground state. Then a Hahn echo pulse sequence is applied~\cite{hahn1950spin}. It involves applying a sequence of pulses on the diamond sample as shown in Fig.~\ref{Fig5_ACMag}. The sequence involves; first a $90^{o}$, then a $180^{o}$ pulse and finally another $90^{o}$ pulse. These pulses are defined by their effect on the spin state of the NV$^{-}$ center when represented on the Bloch sphere. The $m_s = 0$ and  $m_s = 1$ states, conventionally represented as $\ket{0}$ and $\ket{1}$, lie on the poles of the Bloch sphere. The first $90^{o}$ pulse is applied after the NV$^{-}$ center has been prepared in the $m_s = 0$ state. This brings the spin state to the equatorial plane of the Bloch sphere, viz. a coherent superposition of the $m_s = 0$ and $m_s = 1$ states~\cite{hahn1950spin}. The system is then allowed to evolve freely for a time $\tau$; this free evolution of the spin ensemble introduces dispersion (in the equatorial plane of the Bloch sphere) due to magnetic field inhomogeneity. Then a $180^{o}$ pulse is applied to rotate the spin state to the diametrically opposite point on the Bloch sphere equatorial plane ~\cite{hahn1950spin}. Subsequent free evolution, again of duration $\tau$, counters the aforementioned dispersion and refocuses the spin state, which is finally projected onto the $m_s = 0$ and $m_s = 1$ sublevels by the last $90^{0}$ pulse. This is detected in experiments as an `echo' signal that measures the fractional population of the $m_s = 0$ sublevel. This scheme is illustrated in Fig.~\ref{Fig5_ACMag}. Clearly, the effect of any static magnetic field appearing in the first $\tau$ timespan here would be canceled in the second $\tau$ timespan through the inversion of the spin state by the $180^{o}$ pulse. However, for AC magnetic fields, if the time interval between pulses is adjusted such that $2\tau = n.1/\nu$ ($\nu$ is the frequency of the AC field and n is a natural number), then the field magnitude is accumulated in the phase of the echo signal and can be extracted from it~\cite{rondin2014magnetometry}. The experimental realization of this magnetometry technique has been described in \cite{balasubramanian2008nanoscale,maze2008nanoscale,grinolds2013nanoscale}. One of the most important parts of these experiments is to fabricate isotopically pure diamond samples so as to achieve long coherence times. In order to further improve the coherence time of the diamond NV$^{-}$ center, so-called dynamical decoupling techniques are used where the diamond NV$^{-}$ center spin state is decoupled from its environment. These will also be discussed later in the manuscript.
In this technique, the sensitivity can be dramatically improved upto tens of nT for detecting the ac magnetic fields~\cite{degen2008scanning}. If we need to measure a static magnetic field using this technique, the probe itself is made to fluctuate so that the NV$^{-}$ center encounters a fluctuating (ac) magnetic field~\cite{degen2008scanning}. 
In our simulation framework, when a pulse sequence is applied, we need to incorporate the effect of the pulses on the state vector of the system when they are encountered; this is elaborated in Appendix~\ref{Appendix2_PulseSequences}.
The spin sublevels' population is calculated in presence of and spin echo sequence and the ac magnetic field. The population modulation due to AC signal is shown in Fig.~\ref{Fig5_ACMag}. The AC field parameters are taken from~\cite{maze2008nanoscale} sans the environmental noise. Fig.~\ref{Fig5_ACMag} shows the characteristic cosine behavior of the ground state $m_s = 0$ population~\cite{maze2008nanoscale}.

After modeling optical polarization and magnetometry, we now describe the applications of this framework. In order to demonstrate this, we show its usage in designing dynamical decoupling pulse sequences in the next section.

\section{Applications}
\label{Applications}
The framework developed herein could be used to study a range of experimental and theoretical scenarios concerned with the study of diamond NV$^{-}$ center spin dynamics and various sensing application based on it. Here, in order to exhibit the versatility of this framework, we demonstrate one such scenario: the dynamical decoupling of NV$^{-}$ center spin from environmental noise. Dynamic decoupling (DD) is a widely popular technique in NMR. It is used to decouple the nuclear spins under investigation from the environmental perturbations. DD techniques have also found applications in diamond NV$^{-}$ center based high precision magnetometry. DD techniques are used to enhance the decoherence time of the NV$^{-}$ spin by decoupling the environmental degrees of freedom from the spin system so as to improve the sensitivity of the overall system. The best thing about the DD techniques is that the environment need not be controlled in order to decouple it from the system~\cite{souza2012robust}. Depending on the nature of system-environment interactions, various schemes are used for decoupling the system of interest from the environment. One of the most widely used dynamical decoupling technique is Carr-Purcell-Meiboom-Gill (CPMG) pulse sequence~\cite{carr1954effects,meiboom1958modified}. CPMG pulse sequence is able to handle both spin-dephasing and spin-flip noise, though with certain limitations~\cite{souza2012robust}.

\begin{figure}[htbp]
 \centering
\includegraphics[width=8.5cm,height=8.5cm,keepaspectratio]{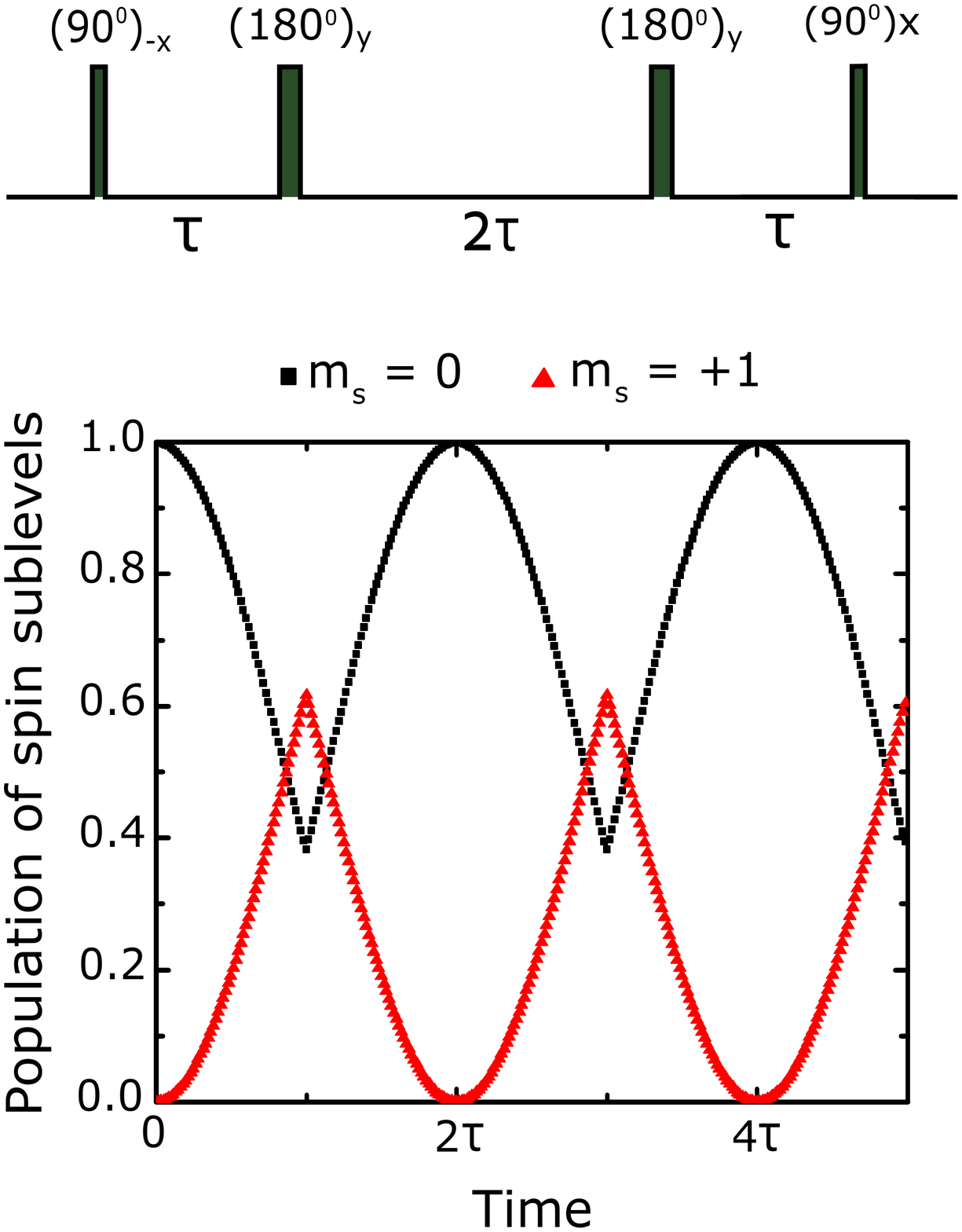}
      \caption{The pulse sequence and ground state spin sublevels' population with CPMG-2 pulse sequence. The subscripts on the pulse labels denote the axis of rotation of magnetic field vector on the Bloch sphere. The initial $90^{\circ}$ pulse brings the spin vector to the y-axis on Bloch sphere and subsequent $180^{\circ}$ pulses flip the spin vector about y-axis eliminating any noise accumulated and resulting in repeated revival of the spin signal in the form of echo at $2\tau$ and $4\tau$.}
      \label{Fig6_CPMG}
\end{figure}
The CPMG technique involves the time reversal of evolution, like Hahn echo sequence, to nullify the deleterious effect of the environment and to reduce the decay rate of the spin vector in transverse plane~\cite{souza2012robust}. The pulse sequence is demonstrated in Fig.~\ref{Fig5_ACMag}. It essentially involves rotation around two axes. The NV$^{-}$ center is initially polarized in $m_s = 0$ state. We consider the dynamics between $m_s = +1$ and $m_s = 0$ state on Bloch sphere. In the CPMG pulse sequence, the first $(90^{\circ})_{-x}$ pulse brings the spin vector to y-axis. Here subscript $-x$ denotes the axis of rotation of the magnetic field vector on the Bloch sphere. Similarly other subscripts have been used for denoting the rotation axes of the pulses. After the initial $(90^{\circ})_{-x}$ pulse, the system evolves freely for a timespan of $\tau$. Then a $(180^{\circ})_y$ pulse is applied. It essentially rotates the spin vector around the y-axis of Bloch sphere by $180$ degrees. Before the $(180^{\circ})_y$ pulse, the spin vector precess around z-axis and spreads in x-y plane, hence oscillation in the population of the spin states. However, it also accumulates the environmental noise signal. The $(180^{\circ})_y$ pulse rotates the spin vectors by $180^{\circ}$ around y-axis, thus reversing the spin precession resulting in signal revival (echo) at time 2$\tau$. The same process repeats after next 180$^{\circ}$ pulses. The repeated application of 180$^{\circ}$ pulses cancels the accumulated environmental noise and gives the repeated echo signal. Depending on the number of 180$^{\circ}$ pulses, the CPMG pulse is also numbered i.e. the CPMG pulse sequence with n 180$^{\circ}$ pulses is called CPMG-n pulse sequence. Fig.~\ref{Fig6_CPMG} demonstrates the CPMG-2 pulse sequences being applied to the NV$^{-}$ center spin state. The characteristic revival of the spin state at 2$\tau$ and 4$\tau$ times exhibits the accuracy of the simulation framework for CPMG sequences. The scheme can be extended for implementing CPMG-n for any value of n.

In addition to the CPMG pulse based dynamical decoupling technique, the proposed model can also model other popular and more general dynamical decoupling techniques such as XY-4 pulse sequence \cite{maudsley1986modified}. These will just be a natural generalization of the demonstrated CPMG pulse sequence based dynamical decoupling technique. Further, dynamical decoupling has become an important ingredient in quantum information processing, spectroscopy and imaging~\cite{biercuk2009optimized,ryan2010robust,souza2011robust,naydenov2011dynamical,bylander2011noise,taylor2008high,de2011single}. Implementation of various kind of dynamical decoupling scheme using diamond NV$^{-}$ center also proves that, along with other more involved dynamical decoupling techniques, various kind of magnetometry or electrometry techniques (e.g. Ramsey magnetometry) can be implemented by the framework presented herein as they also involve application of pulse sequences to extract information about the magnetic or electric field. 

Additionally, nowadays the diamond NV$^{-}$ center is being explored for  probing bio-molecular spin dynamics~\cite{liu2016scheme} and in-vivo cell imaging~\cite{mcguinness2011quantum,perunicic2016quantum}. These application essentially are based on the manipulation of spin states of the NV$^{-}$ center as in the CPMG sequence. Therefore, the simulation framework presented here could be useful for simulating the diamond NV$^{-}$ center based experiments.

\section{Conclusion}
We have presented a quantum biology inspired simulation framework for the diamond NV$^{-}$ center and shown that it successfully models optical polarization, DC magnetometry, AC magnetometry and dynamical decoupling experiments in this system. This framework can enable NV$^{-}$ center based device design. More broadly, it offers a platform to simulate few-spin open quantum systems that are essential for various nanotechnological applications like quantum computation and high precision metrology. 

\appendices
\section{The quantum master equation}
\label{Appendix1_QME}
The Lindblad master equation models the dynamics of an open quantum system i.e. a quantum system coupled to an environment. It models the effect of environmental interaction using operators defined on the Hilbert space of the system itself~\cite{breuer2002theory}. The master equation presented in the text is very much similar to the Lindblad master equation~\cite{breuer2002theory}. It reads:
\begin{eqnarray}
\label{Appendix_ME}
\dot \rho = - \frac{i}{\hbar}[H, \rho] + \sum\limits_{i=1}^{19} k_i (P_i \rho P_i^\dagger - \frac{1}{2}(P_i^\dagger P_i \rho + \rho P_i^\dagger P_i))
\end{eqnarray}
There are two important steps in the construction of this equation. The first one is the construction of the joint Hamiltonian of the ground and excited states along with the intermediate singlet states. This has been depicted in Fig.~\ref{Fig2_JointHamiltonian}. The Hilbert spaces of ground and excited states are independent of each other and the unitary dynamics happens separately in these Hilbert spaces. However, once the NV$^{-}$ center is in excited state (after laser excitation), there are radiative and non-radiative transitions from the excited state to the ground state~\cite{doherty2013nitrogen}. Some of these transitions are direct transitions from excited to ground and some of them go through intermediate singlet states (cf. Fig.~\ref{Fig1_EnergyLevels}). Now the second step in modeling the overall dynamics of the center is to model these radiative and non-radiative transitions. We use projection operators to model these transitions as depicted in Fig.~\ref{Fig2_JointHamiltonian}. The second term on the right hand side of Eq.~\ref{Appendix_ME} models these transitions with the help of projection operators. In a simulation, once the master equation reaches the steady state, we need to calculate the population of the spin sublevels of the ground state. This is achieved by first taking the partial trace of the density matrix and then calculating the expectation value of $m_s = 0$ or $m_s = \pm1$ sublevels of the ground state.

\section{Accounting for pulse sequences in the system state evolution}
\label{Appendix2_PulseSequences}
When we are using pulse sequences (e.g. in the case of ac magnetometry or dynamical decoupling techniques), we take into account the effect of these pulses at the instants when they appear during the evolution. The effect of a $90^{o}$ pulse is to bring the magnetic field vector from the z-axis (pole of Bloch sphere) to the xy (equatorial) plane and vice versa. A $180^{o}$ pulse inverts the magnetic field vector i.e. it converts $\ket{0}$ to $\ket{1}$ on Bloch sphere and vice versa. We can write the matrices corresponding to these pulses, illustrating their effect on the magnetic vectors on the Bloch sphere~\cite{keeler2011understanding}. When using pulse sequences in a simulation, we need to let the system evolve according to its Hamiltonian till a pulse is encountered. And at the moment when the pulse needs to be applied, we need to apply the pulse matrix on the state of the system and afterwards, again, we need to let the system evolve according to its own Hamiltonian until it encounters another pulse. We need to reiterate the same procedure when any other pulse is encountered.

\section*{Acknowledgment}
We are grateful for the support from the Ministry of Electronics and Information Technology through the Centre of Excellence in Nanoelectronics at IIT Bombay.

\ifCLASSOPTIONcaptionsoff
  \newpage
\fi

\bibliographystyle{IEEEtran}
\bibliography{DiamondNVCenter}


%

%







\end{document}